# Business Dynamics in KPI Space

Some thoughts on how business analytics can benefit from using principles of classical physics

*Alex Ushveridze[1]*

*Capella University, Minneapolis, USA*

## Abstract

*The main problem with the methods of machine learning used in today's business analytics is that they do not generalize well and often fail when applied to new data. One of the possible approaches to this problem is to enrich these methods (which almost exclusively are based on statistical algorithms) with some intrinsically deterministic add-ons borrowed from theoretical physics. The idea proposed in this note is to divide the set of Key Performance Indicators (KPIs) characterizing a certain business into the following two distinct groups: 1) highly volatile KPIs mostly determined by external factors and thus poorly controllable by a business, and 2) relatively stable KPIs determined and controlled by a business itself. It looks like, whereas the dynamics of the first group can, as before, be studied by means of statistical methods, for studying and optimizing the dynamics of the second group it is better to use deterministic principles similar to the Principle of Least Action of classical mechanics. Such approach opens a whole bunch of new interesting opportunities in business analytics, with numerous practical applications including diverse aspects of operational and strategic planning, change management, ROI optimization, etc. Uncovering and utilizing dynamical laws of the controllable KPIs would also allow one to use "dynamical invariants" of business as the most natural sets of risk and performance indicators, and facilitate business growth by using effects of "parametric resonance" in natural business cycles.*

## 1. Introduction

### Foreword
The biggest problem any business is facing is the problem of finding practical ways of increasing its effectiveness and creating preconditions for achieving steady growth in the long run. Solving this problem assumes presenting concrete recommendations on what to do on a routine base and how to better plan for future changes.  At first sight, formulating such recommendations for a certain concrete business without knowing in detail every aspect of its internal structure and processes is impossible. However, this assertion is not quite correct.  The point is that whereas the final recommendations themselves should necessarily be formulated in business-specific terms (there are no doubts about that), the methods of their derivation could nevertheless be pretty general and can be formulated with no regard to any concrete business.

---

[1] Email: alex.ushveridze@capella.edu

In this article, I plan to show that the methods allowing such a high level of generality and abstraction may actually exist. They could be based on the very powerful mathematical theory whose roots lie in the area absolutely unrelated to business: this is the area of theoretical physics – or, more exactly – classical mechanics – one of its oldest and best-understood branches. The field of classical mechanics covers literally everything needed for explaining, predicting and planning the behavior of concrete mechanical systems. Its practical applications can be seen everywhere, from computational tasks of mechanical engineering to concrete, practical instructions about the time, direction, duration and power of impulses that should be applied to a rocket, to ensure its safe landing on another planet's surface at a priori specified place and time.

But how the problem of rocket dynamics can be related to the problems the business analysts face on an everyday base? Consider a simple analogy. Imagine that we know how to do predictive modeling but do not know anything about the laws of gravity. Would we be able to send a rocket to a certain planet for the first time? How can we correctly schedule all burns of rocket's engine to ensure its safe landing on planet's surface? This is not a predictive modeling problem in the traditional sense of this word. Each method has its own area of applicability, and predictive modeling is good only if we are able to train the model on large amounts of data more or less accurately reproducing the situation of our interest. But how to train the rocket model if the target has never been reached before?

The situation with businesses is very similar, especially in cases when they try to use analytics for something new, something they had never tried before. How can they go beyond the limits of too poorly generalizable statistical methods? I'm not saying that I know the final answer to this question. But I know that it is very hard (if not impossible) to make a real breakthrough in explaining or predicting the behavior of complex systems without having a simple guiding principle allowing one to look at all their diversity from a single, simple and unified point of view. For rocket dynamics, the role of such a principle is played by the theory of gravity (Newtonian laws). These laws apply to all material objects irrespective of their concrete size and mass, which makes seemingly unrelated and extraordinarily complex problems of space dynamics astonishingly uniform, conceptually simple and mathematically well tractable. But is there anything that could play the role of such a guiding principle for businesses? This is exactly the question I will try to address in this note.

The idea of deriving the laws of microeconomics (i.e. economics of individuals and firms) from general principles is rather old. Irvin Fisher, see e.g. (Fisher, 2006), was probably the first who made a real step forward in this direction more than a century ago. After noticing striking similarities between the concepts of economics and classical mechanics, he conjectured that the right equations of the economic behavior can be obtained by maximizing a certain function of economic variables (the utility function) exactly in the same way as mechanical equations can be derived from the minimization of a function of mechanical variables (classical action). This simple idea has gradually evolved into a broad discipline called the neo-classical theory which continues to dominate in economic studies even today.

Within this theory, the specification of the business economic model assumes two things: selection of independent economic variables, and definition of the so-called utility function – i.e. function that should be optimized. After that, the standard optimization algorithm based on the Lagrange formalism – can be used to derive the needed dynamical equations. The only difference between various approaches is in the ways of choosing the independent variables and the utility function. Traditionally, for independent variables, one uses commodity quantities, while for the utility function -- the raw profit of the business driven by the equilibrium between supply and demand. Let us denote it by $. This choice, being conceptually very intuitive, is however too unsafe from the practical standpoint. The point is that the raw profit is a highly volatile quantity and thus poorly predictable characteristic of business. It strongly depends on the poorly controllable external factors which make its optimization too impractical in the framework of deterministic approaches. This fact has led to the rapid increase of the proportion of statistical methods in economic models and finally resulted in their total prevalence in today's business analytic technologies and tools. The current situation can symbolically be represented as

$$\$ \approx \$_S$$

which means that the currently used approximations of profit function are exclusively based on the use of its statistical representations $\$_S$.

Such a strong disbalance between the usage of statistical and deterministic methods for business needs is something that at least deserves a very careful inspection, especially in the situation when the standard statistical modeling is reaching its natural limits being unable to overcome the difficulties with the generalization. There is a clear gap, and in order to fill it, one should probably look at the deterministic methods again. However to be more successful with this venture we probably should do it in a slightly different way than it was done before. In this paper, I plan to discuss one of such possibilities based on a non-standard selection of independent variables and utility function specifying the business model.

The main idea is to start the process of optimizing the raw profit $ of the business with solving a substantially simpler problem: optimizing its relatively stable component defined as the ability of a business to be profitable. This ability can still be treated as a profit, but as a profit generated under the assumption that all the external factors remain unchanged. It can be measured as a composite value of all those internal assets and processes of a business that are relatively stable and can be controlled by the business itself. In this approach, the role of the independent dynamical variables is played by all the monetizable key performance indicators characterizing these assets and processes. The key point here is just the exclusion of volatile external factors from the consideration. This would allow one to reduce the level of randomness and unpredictability in dynamical processes and focus on only those of their aspects that can be described deterministically. Symbolically, this can be represented as solving an approximate optimization problem for

$$\$ \approx \$_D$$

where $\$_D$ denotes the deterministic part of the profit.

Does this mean that statistical methods should be completely ignored and eliminated from the consideration? Not at all. The idea lies just in the opposite: it is to combine both deterministic and statistical approaches into a single framework. The role of statistics should, however, be changed – in this context it should be used for empowering the deterministic approach, or, in other words, for complementing rather for substituting it. Symbolically, we can represent this combined framework as

$$\$ = \$_D + \$_S,$$

in which the stochastic part of the profit, $\$_S$, should be treated as a correction to the deterministic part and studied by using mathematical methods of stochastically perturbed dynamical systems.

To illustrate what I mean let us consider one more analogy. The designers of car's engine know pretty well that the roads can be bumpy. However, they do not focus on predicting and overcoming each of these random bumps from scratch. Instead, they first try to optimize the internal characteristics of the engine which, being governed by deterministic processes, will make the car more insensitive to external random obstacles of the roads. After solving this main problem, they may fine-tune the parameters of the engine to take into account the statistical properties of the roads too.

The approach I plan to discuss here was first formulated in my previous paper (Ushveridze, 2016) devoted to a broader problem of optimizing the behavior of the autonomous intelligent system in an unknown information space. In that paper the business case was touched very briefly only as a particular illustrative example – simply because the businesses can be treated as very natural examples of such AI systems: they are autonomous, they are intelligent, and their main goal is to survive in an unknown (economic) terrain. Shortly after publishing this paper I came across the papers of Estola, see e.g. (Estola, 2013), who considered similar problems but from purely economic points of view. I realized that his and my approaches have remarkable parallels and belong to one of the 'hot spot' of the business economy. I also realized that making progress in this direction would be critically important for the needs of practical business analytics.

In conclusion of this introductory part let me clearly outline what the reader should and shouldn't expect from the rest of this paper. Its ultimate goal is to try to convince the reader that business analytics as we know it today may have a huge potential for further advancements which can be uncovered if we empower its statistical algorithms by some deterministic methods inspired by classical physics. To make this point maximally clear, I'll intentionally limit myself to discussing the deterministic part of the method only. The problem of its integration with statistical methods of machine learning – which is a key point from the practical standpoint -- will be addressed separately and in a different format. I also will limit myself to discussing exclusively the abstract math and do not plan to touch the business-specific implementation questions at all. I will talk about the businesses not specifying what they actually do and how they make their money. No business applications and no illustrative example based on the use of any real business data will be discussed or provided in this publication.

## Business dynamics

The dynamics of a business can be considered as a motion in its KPI space. KPI stands for the 'Key Performance Indicators' – the parameters characterizing business condition at a given time. As time goes on, these parameters change, and these changes define a certain trajectory in the KPI space. The business usually knows in advance only two points of this trajectory – the starting point representing its current KPIs and the endpoint representing some target KPIs which it wants to have at some future time. If that time is *a priori* known we can say that the business has clear goals.

The hardest problem for any business is just to find the right way of achieving these goals. In the formal mathematical language, this is the problem of choosing a path between two given points in the KPI space. But there are infinitely many paths joining these two points. Which one to select? How to choose the best one? Fortunately, the answer to the last question is easy. For businesses the goodness of a trajectory can be measured in only one way: through the total profit accumulated when moving along it. We can symbolically represent this quantity as $[trajectory] stressing the simple fact that the profit of a

business depends on the way the business moves towards its goals. Acknowledging this fact allows one to ask the next question about the optimal trajectory on which it will be maximal.

It is clear however that in order to answer this question we need to know a little bit more about the possible form of the trajectory->profit relationship encoded in the expression $[trajectory]. Let us see how far we can go in inferring it if we limit ourselves only to very basic information about the businesses and rely exclusively on common sense. First of all, note that, since $[trajectory] has the meaning of the total profit accumulated by a business during its travel time, it would be natural to represent it as a definite integral

$$\$[\text{trajectory}] = \int_{\text{start time}}^{\text{end time}} \frac{d\$[\text{trajectory}(t)]}{dt} dt$$

whose sub-integral expression has the meaning of profit change rate considered as a function of time. At this point we don't make any assumptions – this relation is just an identity.

Now let us try to switch on our common sense. This profit change rate function can obviously be represented as the difference of two parts. One part can be defined as the profit the business may receive through performing some standard, monotonous or, simply speaking, routine operations. The routine itself can be viewed as a dynamical equilibrium established between all processes forming the current business and characterized by a certain location in its KPI space. We denote this part by $P$ which stands for the word '$Potential$'. It is indeed the potential profitability of a business in the situation if external conditions do not change. As to the second part, it represents the losses associated with any attempts to change anything in this routine, or in other words, in the location of a business in KPI space. We denote this part by $C$ which stands for the word '$Change$'. This division is indeed very intuitive because the only way of changing this dynamical equilibrium (which is usually very inertial) is to invest in the changes. Both parts are functions of business KPIs: but while the first part is a function of KPIs themselves, the second one depends on their time-derivatives, i.e. on how fast these KPIs change. This will give us:

$$\int_{\text{start time}}^{\text{end time}} \frac{d\$[\text{trajectory}(t)]}{dt} dt = P\big(\text{location}(t)\big) - C\big(\text{velocity}(t)\big)$$

But if we substitute this expression into the previous one we obtain the expression for the total profit along the trajectory

$$\$[\text{trajectory}] = \int_{\text{start time}}^{\text{end time}} \{P\big(\text{location}(t)\big) - C\big(\text{velocity}(t)\big)\} dt$$

which we want to be maximized.

We have arrived at the mathematical problem having the same form as the least action principle in theoretical mechanics. To make this point clear, let us remember some basic facts about this principle.

## Classical roots

The 'Least action principle' is one of the most fundamental principles of classical mechanics allowing one to derive all the diversity of its motions and dynamical properties from a single and very simple statement (Landau & Lifshitz, 1969; Marion & Thornton, 1988). It states that the problem of describing the motion of a mechanical particle traveling between two points in a coordinate space is equivalent to the problem of finding a trajectory between these two points on which a certain expression @[trajectory] called the 'action' achieves its minimum. The action itself has an extremely simple form and can be represented as an integral over time

$$@[\text{trajectory}] = \int_{\text{start time}}^{\text{end time}} \{C(\text{velocity}(t)) - P(\text{location}(t))\} \, dt$$

As we can see, its sub integral expression is just the difference of two terms. One of these terms represents particle's potential energy which can be qualified as the inactive part of its total energy. It is formed exclusively by the value of the place where the particle is located at time $t$. To have that part of energy the particle does not need to do anything – it just needs to be at that particular place at that particular time. We have denoted this portion of energy by letter $P$ which stands for the word '$Place$'. As to another term, it represents particle's kinetic energy which can be qualified as the active part of its total energy. This energy is determined exclusively by the changes occurring in particles location and thus can be quantified by the velocities of these changes at time $t$. We have denoted this portion of energy by letter $C$ which stands for the word '$Change$'. The difference between the $C$ and $P$ term (or, in more traditional language between the kinetic and potential energies) is usually called the Lagrange function or simply the 'Lagrangian'.

The minimization problem for the action integral turns out to be mathematically well tractable and can be solved by simple mathematical means. Its solution allows us to derive the equations of motion in their most general form and in particular reproduce the famous Newton laws.

Comparing the expressions for action and for profit we see that they are almost identical. There is probably only one seeming difference: here we are looking for the minimum of the action integral while in the business case we looked for the maximum of the profit integral. However, if we note that the sub-integral expression in the business case is $P - C$ while in the mechanical case it is $C - P$, we can conclude that the correspondence between the business and mechanical cases is exact and can be established by the relationship

$$\$[\text{trajectory}] = -@[\text{trajectory}]$$

But this means that the mathematical problem of finding optimal paths for achieving diverse business goals has exactly the same form as the problem of describing the best trajectories for a rocket flying to Mars. This opens an intriguing possibility of importing the rich mathematical formalism of theoretical physics into the area of business analytics for strengthening other more traditional business-focused methods like predictive modeling and data mining.

## Historical remarks

The idea of applying mathematical methods of classical mechanics to the problem of microeconomics (which includes the economics of both individuals and firms) is far from being new. The economists always felt the need for a certain guiding dynamical principle, similar to the principle of least action in classical mechanics, which could be applied to the utility function (Chiarella, 2014). The first step in this direction was made by Irvin Fisher who, in his seminal work (Fisher, 2006) originally published in 1892, conjectured the correspondence between concepts of classical mechanics and economics. In this correspondence, which dominates even in today's economic studies, the particles are treated as individuals (or firms), their coordinates in space as commodity quantities, initial positions as endowments, the velocities as exchanged volumes in commodities, while the energies and their diverse combinations (including the Lagrange function) – as utilities (Elsner, et al., 2015). The existing controversies between various approaches lie in the ways of defining what exactly should be treated as kinetic energy and what as potential.

In early stages of this theory, it was clearly demonstrated that the use of equations of motion allows one to calculate the equilibrium points corresponding to such allocation of commodities in which no mutually beneficial trades are possible, so the balance between the demand and supply is achieved. These points were analogs of the static solutions of equations of motion. The very idea of this theory – that the economics is driven by the equilibrium between demand and supply – turned out to be very instrumental and after inclusion in it the methods of statistics gave rise to the whole branch of modern economics called the neoclassical theory (Mirowski, 1989; Mas-Colell, et al., 1995). The only problem with this theory was related to the range of its applicability, which was essentially limited to the static case. Despite the clear understanding of the fact that this static case is actually an idealization – one of the myriads of more complex and more realistic dynamical behavior, there were no really successful attempts of explicitly describing the character of this dynamics and thus expanding the area of applicability of neoclassical economics.

An interesting attempt to dynamize the static solution was undertaken in (Estola, 2013). By limiting himself to the purely deterministic framework, Estola extended the standard formalism of neoclassical economics by conjecturing possible forms of economic analogs of potential energy. Conceptually, the forms he proposed were very close to those used in the classical mechanics: he defined firm's 'potential energy' as its ability (or, in other words, its potential) to increase its future profitability. However, in contrast to the classical case in which the potential is usually a function of coordinates, he expressed it via the flows of accumulated production – analogs of the velocities in classical mechanics. Nevertheless, using these definitions, Estola was able to develop the dynamical framework for modeling production dynamics of firms. In his framework, the neo-classical theory actually corresponds to the so-called zero-force situation, where the potential energy of the production system achieves its minimum. Estola claims that his approach allowed him to easily describe such difficult cases as firms' permanent growth, business cycles, and bankruptcies (Estola, 2014). These cases couldn't be explained within the standard neo-classical theory where firms are assumed to produce at the constant positive profit maximizing the flow of production.

## Next steps

Beyond any doubt, Estola's approach was a significant step forward in the direction of dynamization of static solutions of neoclassical econometrics. However, I'm absolutely not sure that the direction itself and even the point of departure were chosen correctly. And here is why. In my opinion, the main problem with describing the economics of firms as deterministic dynamical systems is not in finding the appropriate definitions of the analogs of classical potential or kinetic energies but rather in eliminating the randomness factor making their values highly dependent on unpredictable external conditions. This actually was the main reason why the actual progress in solving dynamization problems in the neoclassical model was achieved not within the deterministic approach but rather within statistical methods based on the concept of rational expectations and efficient market view (Mirowski, 1989). But does this mean that the deterministic approach should be forgotten? Absolutely not – quite the contrary -- I even think that it should be considered very seriously because its potential is not fully uncovered. But in order to uncover it, we need some sort of paradigm change: we need to start using dynamics in those domains in which the effects of randomness would be minimal. But how to choose such domains?

The first thing we need to do is to select the right set of basic variables – i.e. coordinates. This set should not include the commodity quantities because their amount depends on the supply/demand balance and thus can be affected by any unpredictable change of external conditions. This actually is a very serious limitation – because it immediately moves us away from the standard neoclassical model in the original Fisher sense. But which variables should be taken as coordinates instead? The answer is simple: any KPIs which characterize the internal assets of the business and are fully controllable by the firm. This may include the number of employees, their skills, the cost of production, the technological base, customer service quality, quality of the product, etc. One requirement to each of these KPIs should be its relative insensitivity to internal noise – for that they should be appropriately aggregated. Other requirements should include the logical and causal independence of chosen KPIs and their relatively small number (the smaller spatial dimension we will work in -- the better).

The second thing we need to do is to select the right functions which would play the roles the potential and kinetic energies in classical mechanics. The analog of potential energy should be a certain function of all the internally controllable KPIs. The main requirement to its form is that it should represent some monetizable quantity – having the meaning of a combined value of all the selected KPIs characterizing a certain stable state of business. As to the analog of kinetic energy, it should have the meaning of the losses associated with intentional KPI changes and, for that reason, should be represented as a function of their time-derivatives.

The high-level sketch of the above program is what I plan to present in the following sections.

## 1. The formalism

### How to define profit?

Denote by $x = \{x_1, x_2, \ldots, x_N\}$ the vector of key monetizable assets of a certain business. The examples of such assets – forming the components of vector $x$ – may include (but not limited to) business' intellectual property and knowhow, IT technologies and level of automatization, infrastructure and mobility, organizational and production effectiveness, quality of sales and marketing, as well as business property, cash, facilities etc. This list can be expanded by diverse quantitative and qualitative aspects of human potential, products, services, compared to industry standards. These assets are directly linked to business performance and thus reflect its current financial health. Using the standard business terminology we hereafter will call these parameters the Key Performance Indicators (or simply KPIs).

If KPIs of a certain business do not change over time, and if there are no uncontrollable external or internal changes (we assume the economic situation is stable and there are no unexpected damaging disasters), then the profit flow of such a business should be a constant which can be explicitly calculated if all KPIs are given. Note that the explicit calculability – which in the mathematical language simply means the existence of a certain function $U[x]$ mapping the set of KPIs onto the profit flow values – is the main requirement to the completeness of KPI set: if for some reason the KPI vector we plan to use is not sufficient to estimate the profit flow, we should supplement it by some missing components allowing us to do so. So, hereafter, by $x$ we will always mean the complete set of KPIs.

Formally, if $x$ is a constant then the profit $d\$$ generated by a business in a certain standard but meaningfully small time interval $dt$ can be represented as a certain computable function of $x$, with the dimension $[d\$/dt]$. Using the profit flow function $U[x]$ we introduced above, we can write:

$$d\$ = U[x]dt.$$

Of course, the stationary case we just described is an idealization and never occurs in reality. Therefore it is tempting to generalize the above formula to non-stationary cases too. The naïve way of doing that would be based on simply replacing the constants $x$ by their time-dependent counterparts $x(t)$. However such a replacement would give us an incorrect estimate of a profit in the non-stationary case. The point is that the quantity $d\$_+$ defined in such a way, i.e. as

$$d\$_+ = U[x(t)]dt,$$

will not represent the profit anymore, but instead only some approximation to it. In fact, this always will be a certain ceiling-type approximation (i.e. the approximation from above) because in case of changing KPIs the number $d\$_+$ turns out to always be larger than the true profit $d\$$. Why? To answer this question it is sufficient to look at who is the initiator of all the KPI changes reflected in function $x(t)$. If, according to our assumption, the external conditions are stable and there are no uncontrollable events, then it literally means that the only candidate for an initiator role is the business itself. But we know that all controlled and conscious internal changes are always associated with losses – simply because any change is the result of a certain planned action and any such action requires some investment to be performed. For example, the business may want to change the product line, increase the quality of services, attract new customers or hire new employees. All these actions will unavoidably result in losses.

But how to quantify these losses? In other words, how to build another term $d\$_-$, which, being subtracted from $d\$_+$, would give us a true profit $d\$$ ? Fortunately, to answer this question we do not need to be too specific. Even very general reasoning may help.

Indeed, no matter which parameters we consider we can formally describe these changes as differences

$$dx(t) = x(t + dt) - x(t)$$

But what actually matters is not the absolute value of these differences but rather their value in a given interval of time. This literally means that the faster we want to achieve some goal, the more resources we need to spend. In our case, this means that instead of $dx(t)$ it is better to use the quantity

$$\dot{x}(t) = \frac{dx(t)}{dt}$$

but (for reasonably small time intervals $dt$) this is nothing but the instant velocity at time $t$! Now it becomes clear that the amount of losses at given time $t$ can given by a certain function $K[\dot{x}(t)]$, having the same dimension as $U[x(t)]$, i.e. $[d\$/dt]$. In other words, we can represent the losses during time interval $dt$ as

$$d\$_- = K[\dot{x}(t)]dt$$

The net profit during the time interval $dt$ is thus given by

$$d\$ = d\$_+ - d\$_-$$

or, equivalently, by

$$d\$ = \{U[x(t)] - K[\dot{x}(t)]\}dt$$

The total profit (i.e. all gains minus all losses) during the time between some $t_1$ and $t_2$ is given by the integral

$$\$ = \int_{t_1}^{t_2} \{U[\boldsymbol{x}(t)] - K[\dot{\boldsymbol{x}}(t)]\}\, dt.$$

This is a quite general expression. It shows that the profit a certain business has accumulated during a certain time $t_2 - t_1$ can easily be calculated if the vector-function $\boldsymbol{x}(t)$, which can be treated as a trajectory in the KPI space, is known. Let us ask the following question:

Given the KPI vector $\boldsymbol{x}_1$ at the time $t_1$ (i.e. where business is today) and given the KPI vector $\boldsymbol{x}_2$ at time $t_2$ (i.e. where business wants to be tomorrow), what is the optimal trajectory $\boldsymbol{x}(t)$ in the space of business' KPIs that connects the points $\boldsymbol{x}_1$ and $\boldsymbol{x}_2$

$$\boldsymbol{x}_1 = \boldsymbol{x}(t_1), \quad \boldsymbol{x}_2 = \boldsymbol{x}(t_2)$$

and maximizes the net profit between these them?

### How to optimize changes?

So how to find the path maximizing this integral? This problem can be solved by using the so-called variation method. This method is very intuitive. It starts with the following obvious remark: if the value $\$$ of the above integral achieves its maximum $\$_0$ at a certain trajectory $\boldsymbol{x}(t)$, then, any perturbation of this trajectory

$$\boldsymbol{x}(t) \to \boldsymbol{x}(t) + \varepsilon \boldsymbol{f}(t)$$

not affecting its boundary conditions

$$\boldsymbol{f}(t_1) = \boldsymbol{0}, \quad \boldsymbol{f}(t_2) = \boldsymbol{0}$$

can only reduce the value of $\$$. Substituting the changed trajectory into the integral

$$\$(\varepsilon) = \int_{t_1}^{t_2} \{U[\boldsymbol{x}(t) + \varepsilon \boldsymbol{f}(t)] - [\dot{\boldsymbol{x}}(t) + \varepsilon \dot{\boldsymbol{f}}(t)]\}\, dt.$$

and assuming that the strength of the perturbation given by the parameter $\varepsilon$ is infinitesimally small ($\varepsilon$ tends to zero), we can expand the resulting expression in powers of $\varepsilon$:

$$\$(\varepsilon) = \$_0 + \varepsilon \$_1 + \varepsilon^2 S_2 + \cdots$$

But the assertion that $\$_0$ is a maximum can only be true if

$$\$_1 = 0$$

for all possible perturbations. The explicit expression for $\$_1$ can easily be obtained and reads

$$\$_1 = \int_{t_1}^{t_2} \left[ \frac{d}{dt}\left(\frac{\partial K[\dot{\boldsymbol{x}}(t)]}{\partial \dot{\boldsymbol{x}}(t)}\right) - \frac{\partial U[\boldsymbol{x}(t)]}{\partial \boldsymbol{x}(t)} \right] \boldsymbol{f}(t)\, dt + \left( \frac{\partial K[\dot{\boldsymbol{x}}(t)]}{\partial \dot{\boldsymbol{x}}(t)} \boldsymbol{f}(t) \right)\Bigg|_{t_1}^{t_2}$$

Since the function $\boldsymbol{f}(t)$ is assumed to be arbitrary except the end points of the trajectory where it is expected to be zero, the only case when this integral is zero is if the condition

$$\frac{d}{dt}\left(\frac{\partial K[\dot{x}(t)]}{\partial \dot{x}(t)}\right) = \frac{\partial U[x(t)]}{\partial x(t)}$$

is satisfied. In other words, the trajectory, maximizing the profit integral should satisfy the above equation known as Lagrange equation. It is easy to see that this is the second-order differential equation, and according to the theory of such equations its most general solution $x(t, C_1, C_2)$ depends on two arbitrary constant vectors $C_1, C_2$ of dimension $N$. The values of these constants can be found from the system of $2N$ equations

$$x_1 = x(t_1, C_1, C_2), \quad x_2 = x(t_2, C_1, C_2)$$

for $2N$ unknowns $C_1, C_2$ which finalizes the construction of the unique trajectory maximizing the integral for $. As we see, the full solution of the above problem requires solving one second-order ordinary differential equation and a system of numerical equations. Despite the seeming complexity of this problem its solution can easily be found by standard numerical methods and doesn't create any technical difficulties for the analysts who want to solve them.

## Using business invariants

The above solution can be used for deriving the conservation laws. This follows from the fact that the general solution of the Lagrange equation,

$$x(t) = x(t, C_1, C_2),$$

depends on 2 vector constants. Differentiating this vector equation by $t$ we obtain another vector equation

$$\dot{x}(t) = \dot{x}(t, C_1, C_2),$$

which, together with the first equation can be considered as a system of 2 vector equations for two vector constants. If we could solve this system explicitly, then this would allow us to write down their solutions as

$$C_1 = f_1[x(t), \dot{x}(t), t]$$

$$C_2 = f_2[x(t), \dot{x}(t), t]$$

i.e. express them via combinations of KPI vectors and their derivatives. This gives us $2N$ independent conservation laws. Here we need to note however that the explicit construction of these laws is not always possible, and for finding them one should use some numerical methods. Among these $2N$ theoretically existing invariants there is however at least one which is explicitly constructible and has a very simple form:

$$E = \dot{x}(t)\frac{\partial K[\dot{x}(t)]}{\partial \dot{x}(t)} - K[\dot{x}(t)] + U[x(t)]$$

The proof that it is actually an invariant is elementary and follows directly from the Lagrange equations. Indeed, differentiating the above expression by $t$ and using the Lagrange equation we obtain

$$\frac{dE}{dt} = 0$$

which means that $E$ is a constant. The invariant $E$ is especially important for the business because it describes a very informative combination of its static and dynamic properties. We will call it the power of the business and will discuss in detail in one of the following sections.

In this connection, it is important to note that the Lagrange equation we started with can also be considered as a conserved quantity. Indeed, introducing the vectors defined as

$$\boldsymbol{D} = \frac{d}{dt}\left(\frac{\partial K[\dot{\boldsymbol{x}}(t)]}{\partial \dot{\boldsymbol{x}}(t)}\right) - \left(\frac{\partial U[\boldsymbol{x}(t)]}{\partial \boldsymbol{x}(t)}\right)$$

and thus comprised of $\boldsymbol{x}(t), \dot{\boldsymbol{x}}(t)$ and $\ddot{\boldsymbol{x}}(t)$, we can say that according to Lagrange equations these combinations are always zero. So these quantities can be called super-invariants: they not only do not change over time but are always zero

$$\boldsymbol{D} = \boldsymbol{0}.$$

Now, why would we need to have all these conservation laws? The answer is very simple: because the conserved quantities are the best characteristics of the business. Indeed, what we have essentially showed is that if the trajectory in KPI space is optimal (i.e. leads to that business' goals in the most optimal way maximizing the total profit along the trajectory), then there necessarily should exist some quantities comprised of these KPIs and their derivatives that will remain unchanged no matter how the particular KPIs are changing. And the importance of knowing these invariants is threefold:

- They are the best natural indicators of the closeness of business dynamics to its optimum. Indeed, the fact that they are not stable indicates that business does not evolve in an optimal way. For the same reason, they can be used as early warning signs for the appearance of some undesirable trends. Indeed, if they were constants and then suddenly changed – this means that something went wrong.
- They can be used as most reliable differentiators of business – quantities that can represent its profile better than any other KPIs can. Simply because only the presence of stable features can make the objects better recognizable and uniquely identifiable.
- They make the businesses more predictable and simplifies the operational planning.

The last statement needs special attention.

## Operational planning

What is especially amazing about the business invariants is that they have a real predictive power. And this is because they represent relations between the different instances of time. Consider for example the second-order invariants $\boldsymbol{D}$ which are simply equivalent to equations of motion. In their most general form, these equations express the relationships between the trajectory $\boldsymbol{x}(t)$, its time-derivative $\dot{\boldsymbol{x}}(t)$, and the second time-derivative $\ddot{\boldsymbol{x}}(t)$:

$$\boldsymbol{D}[\boldsymbol{x}(t), \dot{\boldsymbol{x}}(t), \ddot{\boldsymbol{x}}(t)] = \boldsymbol{0}$$

but such things like derivatives are sort of idealizations, especially in a business environment. Therefore it would be much more natural to approximate them in some way. To make things especially clear, we can rewrite these derivatives in the discrete form assuming that the time interval Δt between the moments when we measure KPIs and make some decisions about the next steps is finite. Replacing the first- and second-order time derivatives with their discrete analogs

$$\dot{x}(t) \to \frac{x(t) - x(t - \Delta t)}{\Delta t}$$

$$\ddot{x}(t) \to \frac{x(t + \Delta t) - 2x(t) + x(t - \Delta t)}{(\Delta t)^2}$$

we will see that the above relation between $x(t)$, $\dot{x}(t)$, and $\ddot{x}(t)$ transforms into a certain system of relations between $x(t + \Delta t)$, $x(t)$ and $x(t - \Delta t)$. Resolving this system with respect to $x(t + \Delta t)$ we obtain the relation

$$x(t + \Delta t) = R[x(t), x(t - \Delta t)]$$

whose meaning is more than transparent: it instructs the business what should be its next set of KPIs if its current KPIs and at least one instance of historical ones are known.

At first sight, the first-order invariants, like $C_1$ or $C_2$, seem to be have even higher predictive power. Indeed they connect only $x(t)$ and $\dot{x}(t)$ and their discretized versions can be rewritten as relationships between $x(t)$ and $x(t + \Delta t)$ which would allow the business to estimate the future optimal values of their KPIs based on their current values. No historical data is needed. There is however a price for that increased predictive power, because to correctly estimate the values of future KPIs, the business not only needs to know their current values but also the values of the corresponding invariants.

The prediction of the values of $x(t + \Delta t)$ is what in business language is called operational planning. The relations like we have described above makes this operational planning a routine process. If the goals are set then the optimal way towards that goal breaks down into a sequence of identical routine operations. One does not to think to perform these operations: they are exactly the same all the time: the function $R$ that encodes them do not change: the only thing that actually changes is the data – i.e. the KPIs. If we know their current and previous values, then the function $R$ allows one to compute the next ones. Achieving the latter is what business' operational (or immediate) goals are. But as soon as these goals are achieved, the triple formed of the past, current and next KPIs rotates. The current KPIs become past, next KPIs become current, and the problem arises to find the new next KPIs so the cycle repeats.

But how to determine the form of function $R$ ? To do that we need to know the form of equations of motion. But this, in turn, requires the knowledge of functions $K$ and $U$ that determine the losses and gains.

## 2. Business metrics

### Quantifying losses
It turns out that quantifying losses is relatively easy. Indeed, consider the loss function $K[\dot{x}(t)]$ and try to guess its possible form. First of all, we know that if the function $x(t)$ is constant then there shouldn't be any losses. But if function $x(t)$ is constant, its derivative $\dot{x}(t)$ should be zero. This means that

$$K[\mathbf{0}] = 0$$

Now, remember that we decided to limit ourselves to the cases when all changes in business's KPIs are caused by some internal and apriori planned actions only. In that case, any change of KPI's will

unavoidably lead to some losses resulting in positive values of function $K$. But this means that function $K$ should have its minimum at the point where it is zero. In other words

$$\nabla K[\mathbf{0}] = \mathbf{0}$$

The next assumption is about the magnitude of these changes. It is natural to assume that there are small enough, which means that even if the business makes any changes, it makes them slowly. But assuming this is equivalent to saying that if we expand function $K[\dot{\mathbf{x}}(t)]$ in powers of $\dot{\mathbf{x}}(t)$ it would be accurate enough to keep in this expansion only terms quadratic in $\dot{\mathbf{x}}(t)$. Note that the zeroth-order and first-order terms will be absent because of the two above conditions. This will allow one to write the most realistic expression for losses which will have the following simple form:

$$K[\dot{\mathbf{x}}(t)] = \sum_{i,k} K_{ik} \dot{x}_i(t) \dot{x}_k(t)$$

Here $K_{ik}$ is a certain symmetric and positive definite matrix. The positive definiteness follows from the fact that the loss-terms should always be positive. If the number of KPIs (or the dimension of vector $\mathbf{x}(t)$ is $N$ then the maximal number of independent entries of such a matrix is $N(N-1)/2$. However, it is easy to show that the number of essential independent parameters characterizing the losses is only $N$. To show this, we can play a little bit with the components of the KPI vector $\mathbf{x}(t)$. For example, nothing prevents us from replacing the original components $x_i(t)$ with their linear combinations

$$x_i(t) \rightarrow \sum_k S_{ik} x_k(t)$$

Such a replacement does not change anything with a business – the only thing it changes is the way we describe it. It does not matter are we using for this description the original KPIs or the modified ones. The only thing we need to care of is the equivalence between the old and new sets, which means that both sets should contain the same information (it shouldn't be "lost in translation"). From the mathematical standpoint, this equivalence means the invertibility of matrix $S_{ik}$.

From the general theory of matrices, it follows that this is always possible to choose matrix $S_{ik}$ in such a way that

$$\sum_{i,k} K_{ik} S_{in} S_{kn} = K_n, \quad \text{where } K_n > 0$$

$$\sum_{i,k} K_{ik} S_{in} S_{km} = 0, \quad \text{if } n \neq m$$

This is a direct consequence of the fact that $K_{ik}$ is symmetric and positive-definite matrix. The $N$ numbers $K_n$ are called its eigenvalues of matrix $K_{ik}$. In our case it is natural to call them 'eigenlosses'. In terms of the new, transformed KPIs the loss term takes especially simple form:

$$K[\dot{\mathbf{x}}(t)] \rightarrow \sum_n K_n \dot{x}_n^2(t)$$

The procedure we just described is similar to that used in the principal component analysis (PCA): similar to the later, we replace the original data points with their linear combinations in a way that helps one to make them maximally decorrelated – i.e. independent and separated from each other. We essentially

have demonstrated that such a separability is achievable for any loss function: we can easily identify the 'directions' in the KPI space in which the losses become maximally independent from each other and quantifiable with the minimal set of parameters.

Looking at these numbers from the standpoint of classical mechanics it is natural to interpret them as inertia coefficients, or masses. The key point here is that exactly as in the case of mechanical systems each of the established business processes always has some inertia measured as the level of investments necessary to make the necessary changes in it. This is like the inertia of a rotating wheel. The inertia of the light wheel is small, and it is very easy to accelerate or slow it down. But if the wheel is heavy so its inertia is large too then any changes in its speed may require a considerable investment of energy. The amount of that investment may strongly depend on the character of changes, or if we use our PCA language, on the direction in the KPI space. For example, if an independent contractor will decide to change its hourly rate, this will obviously affect its annual income, but the changes themselves won't cost him any cent. But if the same contractor will decide to change the type of service or sell a new product, this may require learning new skills, getting a license, developing that product, marketing it, which obviously may require quite remarkable investment.

## Quantifying gains

We talked a lot about the form of the loss function $K[\dot{x}(t)]$ and showed that identification of its general form is a relatively easy task in case if its vector argument $\dot{x}(t)$ is small. Based on a rather general assumptions and common sense, we showed that in that case its form is pretty rigid and all its diversity can be reduced to $N$ independent parameters which we dubbed eigenlosses. The problem of finding all these eigenlosses is also a relatively easy task for a business. This is because the business usually has full control of the values of $\dot{x}(t)$ and thus can easily investigate the local neighborhood of the point $\dot{x}(t) = \mathbf{0}$ (zero speed point) to deduce the form of function $K[\dot{x}(t)]$ in that area.

But what about the gain function $U[x(t)]$? Is it constructible in a similar way? Unfortunately not. It turns out that the problem of its definition is much harder and the form of this function strongly depends on business specifics. Finding this function is the main goal of what in business terminology is called 'strategic planning'. The reason making the identification of the gain-function so hard is that in order to really benefit from its knowledge, the business should know it in a global area. Global in the sense that it should include both vectors $x_1$ and $x_2$ (i.e. both current and desired KPIs) as well as the entire trajectory $x(t)$ connecting these two points. Since the latter is not assumed to be known in advance, we essentially need to know $U[x(t)]$ in a much broader area containing a whole bunch of potential trajectories to allow the variation method to select the best of them.

Another circumstance making things especially complex is the fact that the shape of function $U[x(t)]$ may change during the time when the business moves from the point $x_1$ to point $x_2$ in its KPI space. To apply the method we just described it is important to know the form of function $U[x(t)]$ in the vicinity of a certain intermediate point $x_0$ at time when the trajectory $x(t)$ crosses it.

So we see that the determination of function $U[x(t)]$ is intrinsically non-local from both spatial (in the KPI sense) and temporal points of view. Its shape could be highly complicated. While finding its exact form is an absolutely unrealistic task, looking for approximations could be a good alternative. The best way of constructing such approximations is to use sequential approach. Each step of this sequence starts with examining the neighborhood of the point $x_1$ the business currently resides. For that purpose, the business may need to conduct market research to analyze possible forms of function $U[x(t)]$ in the given neighborhood. The goal of this analysis is to be able to select the destination point $x_2$ belonging to the same neighborhood and then solve the problem analytically to find optimal trajectory between the

points $x_1$ and $x_2$. As soon as this is done start moving to the destination along the found trajectory following the routine rules of the equations of motion. The step is completed when business reaches its destination. Then everything repeats indefinitely many times. The success of each step depends on the size of the neighborhood in which the function $U[x(t)]$ can be described reasonably well. The larger the neighborhoods are – the more options the business may have for setting up its goals and finding optimal ways of reaching them. Having larger neighborhoods is important for another reason too: this allows one to reduce the number of market researches the business needs to conduct each time when it sets up a new goal.

We plan to cover this part in a separate publication. Here we want to consider some specific local shapes of function $U[x(t)]$ which the business may encounter when moving along the pass and whose specifics is important to know. To make things simpler and fully focus on studying function $U[x(t)]$ we reformulate the profit integral in terms of transformed variables $z_i$ and then absorb the eigenlosses $K_i$ in these variables by rescaling them as

$$x_i \to \frac{x_i}{\sqrt{2K_i}}$$

This makes the loss term expressed in terms of new variables trivial:

$$K[\dot{x}(t)] \to \frac{1}{2}\sum_n \dot{x}_n^2(t) = \frac{1}{2}\dot{x}^2(t)$$

and all specifics of the system becomes concentrated in the gain term. This will result in the following form of the profit integral

$$\$ = \int_{t_1}^{t_2} \left\{ U[x(t)] - \frac{1}{2}\dot{x}^2(t) \right\} dt$$

and also to the master equations describing the optimal path

$$m\ddot{x}(t) = -\nabla U[x(t)]$$

and the quantity

$$E = U[x(t)] + \frac{1}{2}\dot{x}^2(t)$$

which is conserved if the business moves along the path.

## Quantifying power

It is a good time to discuss the power of the business – a conserved quantity which has been introduced above and denoted by $E$. The form of the loss term we have just derived allows one to simplify the original expression for $E$. Indeed, since $K[\dot{x}(t)]$ is quadratic in $\dot{x}(t)$, the following relation holds:

$$\dot{x}(t) \frac{\partial K[\dot{x}(t)]}{\partial \dot{x}(t)} = 2K[\dot{x}(t)]$$

From this relation and definition of $E$ it immediately follows a very simple expression for $E$:

$$E = U[x(t)] + K[\dot{x}(t)]$$

But let us ask the following question: is there indeed a good reason for adding the gains and losses? What is the meaning of such a strange combination? What we actually use is their difference: it is very natural because it gives us the profit. But what can give us their sum?

To answer this question, we probably first should stress the fact that this sum is conserved only if the trajectory satisfies the equations of motion which, in turn, means that it is optimal. Keeping this fact in mind, let us try to articulate explicitly what conservation of this sum actually means. It literally means that if the gains given by the terms $U[x(t)]$ are small then one should increase the loss terms $K[\dot{x}(t)]$ to keep their sum unchanged. This is what system treats as optimal behavior! But isn't it counterintuitive? Indeed, if we increase the losses at the time $t$ when our gains are low we simply double our losses. Is this behavior optimal?

The whole point here is that this seeming non-optimality occurs only if we limit our analysis to some specific instants of time $t$. At some other instances of time the situation may just switch to totally opposite: large gains and small losses, for example. This may happen all the time because the system is not designed to optimize the instantaneous profit. It tries to optimize the integral profit instead! To do so, it looks far ahead and, after analyzing the profile of the landscape between its start and end points, comes to conclusion that if the current KPIs are not very favorable for the business, the best business strategy would be to change these KPIs to something better. Very logical, isn't it? But the only way of doing this is to invest in these changes, and this investment is always a synonym of losses! It is hard to disagree with this explanation which actually states that to have global gains in the future it is necessary to invest in this future today – i.e. be ready to admit some temporary losses too.

## 3. Business dynamics

### Static traps

The most interesting type of the gain function $U[x(t)]$ appears if it has a local minimum at a certain point $x_0$ in the modified KPI space. Without loss of generality we can assume that this point lies in zero: $x_0 = 0$, because, otherwise, we can easily redefine the parameters as $x(t) \to x(t) + x_0$ and thus reduce the problem to the case when the minimum is attained at $x_0 = 0$. Let $U_0$ denote the value of this minimum

$$U[0] = U_0$$

Then, taking into account that

$$\nabla U[0] = 0$$

we can expand the function $U[x(t)]$ around this point and assuming that the deviations from the minimum are small, write the expression

$$U[x(t)] = U_0 + \frac{1}{2}\sum_{i,k} U_{ik} x_i(t) x_k(t)$$

which is quadratic in vector components, contains a symmetric and positive definite matrix, and thus looks very similar to the analogous expression we have derived for the loss term. Here we probably need to stress again the principal difference between these two expressions. Despite the fact that the expression for the losses looks simple, it is quite general because its local minimum at zero is at the same time the global minimum too and therefore, even considering small deviations from it, we are actually getting insights about the general picture. As to the expression for the gains, it, as we have already noted above, may have (and usually has) multiple local minima, which prevents us from making any conclusions about the global picture if we restrict ourselves to only studying how the small deviations from each of them look like. Nevertheless, these local minima have a very important meaning: they are sort of traps and getting out from them could be a big challenge for any business. Let us explain why.

We have already noted that the best strategy for any business is to follow the equations of motion in the KPI space. However, even this best strategy may not suffice in cases when the business has insufficient power and resides in one of such local minima. In that case, one of the possible strategies the business may select is to not do anything except maintaining the existent status quo. This, by the way, is a quite viable strategy, totally satisfying the equations of motion, because not doing anything ($m\ddot{x}(t) = 0$) and being in the minimum ($\nabla U[x(t)] = 0$) means that both left and right hand sides of the equations of motion are zero. In that case, the power of the business is determined exclusively by the constant term in the expansion for gains:

$$E = U_0$$

which can be interpreted as a constant income the business may have – and whose source could be the interest of some equities it owns, for example. If in that case, the business decides to move out of the local minimum, it will need to invest in some changes, but both the initial investment flow and the deviation from the minimum will be limited by this value of $E$. This means that if the power of a business is initially low, it may not be able to set big goals – or, in other words, get out of the wells it currently resides in which it can function indefinitely long time.

Is there any way of getting out of such traps – a way which would be justified scientifically and would allow practical implementation without considerable investments? It seems that the answer is yes and the idea of such a way is to use a trick originating from the theory of oscillators developed in theoretical mechanics. But in order to apply this trick, we first need to explain why the oscillators may have something in common with businesses.

### Stationary eigen-cycles

To further simplify this expression let us manipulate again with the set of KPI components $y_i(t)$. We have already simplified the quadratic expression for the loss term. Exactly the same procedure can be applied to the quadratic gain term too. Let us consider the transformation

$$x_i(t) \rightarrow \sum_k S_{ik} x_k(t)$$

with some unknown coefficients $S_{ik}$ forming a $N \times N$ matrix. If this matrix is orthogonal then, this transform will not affect the form of the loss term.

As to the gain term, it will obviously change. Omitting the details, we can claim that there always exists such an orthogonal transform $S_{ik}$ that diagonalizes the matrix $U_{ik}$ and thus brings the gain term to the canonical form:

$$U[s(t)] = U_0 + \frac{1}{2}\sum_i \omega_i^2 x_i^2(t)$$

Here the numbers $\omega_i$ are the so-called eigenfrequencies which can be easily constructed in a purely algebraic way if the matrix $V_{ik}$ is given. The term 'eigenfrequency' is a combination of two words 'eigen' and 'frequency'. The first one reflects the fact that the numbers $\omega_i$ appear as solutions of the eigenvalue problem for matrices $V_{ik}$. The second one stresses the fact that these numbers are actual frequencies the system under consideration oscillates with. The last statement should be not too surprising for those who are familiar with the Lagrange function for the system of simple harmonic oscillators: we just obtained this function. However, we do not need to know anything about mechanics of oscillators to make sure that the statement is correct. We just need to find the optimal trajectory for that Lagrangian and look at it. Fortunately, this is not difficult, because in terms of new variables, the Lagrange equations become especially simple and read:

$$\ddot{x}_i(t) + \omega_i^2 x_i(t) = 0, \quad i = 1,\ldots,N$$

First of all, we see that they are completely separated, so we actually deal with multiple copies of a single equation, each with its own parameter $\omega_i$. Second, all these equations are immediately solvable, and their explicit solutions have the form:

$$x_i(t) = a_i \sin \omega_i (t - c_i)$$

Here $a_i$ and $c_i$ are two arbitrary constants which can be established from the initial conditions. The initial conditions in that case read:

$$x_i(t_1) = x_{i1}, \quad x_i(t_2) = x_{i2}, \quad i = 1,\ldots,N$$

and reduce to the system of $2N$ equations:

$$a_i \sin \omega_i (t_1 - c_i) = x_{i1}, \quad a_i \sin \omega_i (t_2 - c_i) = x_{i2}, \quad i = 1,\ldots,N$$

which can be used for finding $a_i$ and $c_i$.

The appearance of frequencies which we called 'eigenfrequencies' is one of the direct consequences of the approach we discuss here. These eigenfrequencies are not something unnatural to the business – they reflect a very core property of any business dynamics – its intrinsic cyclicity. The fact that we obtained a whole spectrum of such frequencies is also very typical for most of the businesses – usually, they are characterized by many different coexisting periods or cycles. These cycles may include financial cycles, reporting cycles, service cycles, product development cycles, etc. Many of these business process-specific cycles are aligned to the natural cycles, like, for example, daylight hours, work weeks, seasons, years, etc. The reason for such an overall cyclicity is the fact that by organizing work in cycles is the most optimal way of saving business resources. The periodic motion is always more economical than a random walk.

However, an obvious drawback of cyclicity is that the processes revealing this property never end. This means that the analogy with business traps can be extended. These traps can be dynamical too. There is actually nothing wrong with that if we restrict our theory to the case that the external situation doesn't

change. In that case, the business may be in a stable dynamical equilibrium with the environment and reside in such a state indefinitely long time. However, such a situation is obviously an idealization.

## Towards non-stationarity

As we said above the stationary situation is an idealization. On the general level described by the general gain and loss terms, the stationarity condition means that these terms do not explicitly depend on time so that all their dependence on time occurs only via their time-dependent arguments: the KPI vectors and their derivatives. If we consider the specific case we just discussed in the last section – the stationarity means that the matrices are time-independent. However, as we said, this is an idealization that never occurs in reality. What does it change in our formalism? Nothing – if we consider it on the conceptual level. However it may affect the form of equations of motion and some of the business invariants: for example, the business power will not be an invariant anymore, unless we appropriately modify its form. Even in an oversimplified case when there is only one KPI parameter which together with its derivative is being kept small so that the resulting equations of motion are one-dimensional and linear, their form, compared with the stationary case:

$$\ddot{x}_i(t) + \omega_i^2 x_i(t) = 0$$

will become substantially more complex

$$\ddot{x}_i(t) + \sum_k q_{ik}(t)\dot{x}_k(t) + \sum_k k_{ik}(t)x_k(t) + f_i(t) = 0$$

The first source of complexity comes from the fact that the coefficients of this equation become functions of time. And the main point here is that these functions are not under business' control. Otherwise, we would treat them as additional business KPIs. The second source of complexity is that we see here some new terms which have not been present in the stationary case.

We can simplify the analysis of this time-dependent equation if we assume that it is a small perturbation of the stationary one. Saying this is equivalent to requiring that all the newly introduced terms are small: i.e. $q_{ik}(t) \to 0$ and $f_i(t) \to 0$, and the modifications of the existing terms are also small: i.e. $k_{ii}(t) = \omega_i^2 + p_{ii}(t)$, where $p_{ii}(t) \to 0$, and $k_{ik}(t) = p_{ik}(t)$, where $p_{ik}(t) \to 0$ if $i \neq k$. This gives us

$$\ddot{x}_i(t) + \omega_i^2 x_i(t) = F_i(t)$$

where

$$F_i(t) = -\sum_k q_{ik}(t)\dot{x}_k(t) - \sum_k p_{ik}(t)x_k(t) - f_i(t)$$

In other words, we get the oscillator in the external force field. We see that the external forces consist of three terms. One is independent on KPIs, second – dependent on KPIs themselves, and the third one – dependent on KPIs derivatives, i.e. on the rates of their change. Let us consider now how does each of these terms affect the behavior of oscillator. One thing we can say with confidence before conducting any analysis is that the power of the business defined by the old expression

$$E(t) = \frac{1}{2}\sum_i \dot{x}_i^2(t) + \frac{1}{2}\sum_i \omega_i^2 x_i^2(t)$$

will not be a conserved quantity anymore. Depending on the character of these three components of external forces, it will either increase or decrease. Some of their combinations will lead to the increase of the business power, we will call them good or constructive forces, and some will decrease it, we call them bad or destructive forces. It is very important to understand the nature of these destructive or constructive forces. The main question which we will be interested in is the question about concrete criteria that would allow one to easily separate good forces from the bad ones and thus make the behavior of business more predictable. The goal of every business is to be able to timely recognize any change in the character of external forces in order to start consciously manipulating them – i.e. either avoiding them or using for business needs, depending on the situation. Below we show how the theory of oscillators allows one to easily answer that question. I do not intend to go into details of all possible scenarios which, being easily analyzable, are rather cumbersome to be described in terms of simple math we used so far in this paper. It would be much more instructive to focus on the most interesting cases.

## How to facilitate business growth?

Let us start with the case when the external forces are caused exclusively by the term that does not depend on the KPIs:

$$F_i(t) = -f_i(t)$$

This case can best be explained if we consider swing. We know very well that if we slightly push swing in the direction of motion each time when it crosses its minimum, and so has the maximal speed, then we can gradually increase its amplitude and thus its energy as well. This effect is known as resonance. Translated into the language of business it would mean that if there are some external circumstances that arise with the periodicity coinciding with business' eigenfrequencies

$$f_i(t) = a_i \sin(\omega_i t + \varphi_i)$$

and, by a fortune coincidence, their effect has the 'right' direction each time when they appear, then they may dramatically increase business power. At first sight, this seems to be a highly constructive effect, but in reality, it, unfortunately, has a serious drawback. It is too risky to rely on exclusively external forces hoping that they will be favorable to the business. The chances for accidental coincidences of external and internal frequencies and the coherences of their phases are usually negligibly small.

A much more promising case appears when the external term does explicitly depend on KPIs and has, for example, the form

$$F_i(t) = -p_i(t)x_i(t)$$

We intentionally avoid considering the most general form of this term in order to get closer to the point. In that case, the business changes its internal parameters in a way that effectively results in the change of its eigenfrequencies. Let us consider an especially interesting case when

$$p_i(t) = a_i \sin(2\omega_i t + \varphi_i)$$

i.e. when the frequencies with which the parameters $p_i(t)$ oscillate are twice larger than the corresponding business eigenfrequencies. To see why such exotic frequencies might be interesting to us, let us note that since the unperturbed solution of the dynamical equation has the form

$$x_i(t) = x_i \sin(\omega_i t + \theta_i)$$

the external term $F_i(t)$ describing the perturbation can approximately be represented as

$$F_i(t) = -x_i a_i \sin(2\omega_i t + \varphi_i) \sin(\omega_i t + \theta_i)$$

But, according to well-known trigonometric formulas, this term will reduce to the sum of two oscillating terms

$$F_i(t) = x_i a_i \cos(\omega_i t + \varphi_i - \theta_i) - x_i a_i \cos(3\omega_i t + \varphi_i + \theta_i)$$

the frequency of one of which coincides with the corresponding business eigenfrequency while the frequency of another one is three times larger. But it is obvious that the first term will cause resonance!

The main difference of this resonance from the previous one we considered above is that it can be caused by internal changes. As we can easily see, the effect of introduction of the 'external' term we just considered is exactly the same as if we would replace the original constant eigenfrequencies $\omega_i$ with their 'slightly oscillating' versions as

$$\omega_i^2 \to \omega_i^2 - a_i \sin(2\omega_i t + \varphi_i)$$

But since the eigenfrequencies are certain functions of KPIs, the same effect can be achieved if we start slightly perturbing the KPIs with small oscillations whose periods are exactly two times shorter than the periods of basic business cycles. This effect in classical mechanics is called the parametric resonance. We see how to use it in businesses to facilitate their growth.

In conclusion, let us briefly mention the third case when the external term depends on KPI change rate:

$$F_i(t) = -q_i(t)\dot{x}_i(t)$$

This is the case of the so-called damped harmonic oscillator, whose behavior qualitatively depends on the sign of the components of function $q_i(t)$. If these components are positive, then the external term imitates friction-like forces whose direction is opposite to the direction of motion. The oscillator is gradually slowing down, its amplitude decreases and it eventually stops. In business terms it means that business gets trapped in one of the local minima of the global gain function. So we have in that case a clearly destructive behavior. However, if the functions $q_i(t)$ are negative, then the external term starts working as an accelerator, the amplitude of oscillations gradually increases and business eventually gets out of the trap. This case may seem highly favorable for the businesses, especially because there is no need for pumping the KPI parameters with double frequencies – the accelerating effect can easily be achieved with constant functions

$$q_i(t) = q_i$$

However, a more detailed mathematical analysis shows that it is very difficult (practically impossible) to reproduce the constant anti-dissipative term in the framework of Lagrange formalism which makes this case essentially non-quantifiable in terms of optimal profit-generating dynamics.

## Afterword
Businesses have astonishingly many features of both intelligent, living, and non-living (i.e. purely mechanical) systems (Ushveridze, 2016). Indeed, if we look at any business as biologists, it will appear to us as an extremely complex living organism facing the problem of survival in an unfriendly (economic) terrain and forced to constantly search for (financial) resources. We will see that many facts about that business, including its internal organization and external behavior, can easily be formulated and understood in a purely biological language. Psychologists (as well as computer scientists working in the area of AI) will probably find in the same business many similarities with intelligent systems: the ability to learn and interpret things, analyze diverse situations and adequately react to them, plan, negotiate, make smart decisions or stupid mistakes and behave either rationally or irrationally depending on the situation. As to the physicists – they may prefer to treat this business as a mechanical system and explain its behavior in terms of conservation laws, optimization principles and equations of motion. The list of such analogies can be continued further, and this should not surprise us too much because nature is known to be rather uniform and patterns that seem specific to one area can easily be found in different and seemingly unrelated areas too. But the central question is how can we benefit from this uniformity? This is probably something that every business owner may want to know at the first place after hearing about all these similarities and analogies: how they may help in solving very concrete business problems?

In this paper, I made an attempt of approaching this question from the standpoint of theoretical physics by studying the analogy between the businesses and mechanical systems. In contrast with other two analogies (I mean the living and intelligent systems), the classical mechanics has a great advantage – it has enormous mathematical support making it especially powerful for practical applications. I tried to utilize this fact and showed that by using its simple and yet very powerful mathematical formalism, it is possible not only to better understand many aspects of business dynamics but also formulate concrete recommendations on how to increase its effectiveness.

However, one should make it very clear here that the form in which we have exposed the material of this paper was an idealization. Our goal was to demonstrate the power of the main idea by completely obscuring the details of its possible implementation. There are at least two aspects that should be taken into account when trying to use all the above considerations in practice.

1. We have implicitly assumed that the functions describing business environment are analytic functions, i.e. they are continuous, differentiable, expandable into power series, etc. We freely operated with such notions as velocity, being perfectly aware of the impossibility of using these approximations in the case of real data. We have illustrated our exposition on the artificial and oversimplified examples of linear equations allowing explicit and simple-looking solutions. In a real situation, it is rarely the case.
2. We limited ourselves to describing business environment in the language of a fully deterministic theory. But we know very well that such entities as businesses are not deterministic objects, and any of their descriptions that do not take into account the randomness factor will be inherently incomplete. The right approach is to integrate the ideas of the deterministic Lagrange approach into the statistical methods of machine learning. This integration can be performed in a very natural way, and this is one of the subjects I plan to cover in the near future.

In conclusion, I would like to stress that the method exposed in this paper is not limited to the businesses considered as a whole. It is equally well applicable to their diverse parts, like for example departments, product lines, or even individual customers.